\documentclass[prl]{revtex4}
\usepackage{amsmath,amssymb,latexsym,epsfig,graphics,epsf}
\usepackage{graphics}
\usepackage{float}
\newcommand{\br}{{\mathbf r}}

\newcommand{\tr}{{\tilde{r}}}
\newcommand{\tv}{{\tilde{v}}}

\newcommand{\tbF}{{\tilde{\mathbf F}}}
\newcommand{\bR}{{\mathbf R}}
\newcommand{\tbR}{{\tilde{\mathbf R}}}

\newcommand{\thf}{\tilde h}
\newcommand{\tgf}{\tilde g}

\newcommand{\tU}{\tilde{\Phi}}
\newcommand{\tUe}{\tilde{\Phi}_{{\rm  EXP}}}

\newcommand{\be}{\begin{equation}}
\newcommand{\ee}{\end{equation}}

\newcommand{\esub}{{\mbox{\tiny EXP}}}
\newcommand{\e}{{\rm  EXP}}
\begin{document}

\title{Explaining why simple liquids are quasi-universal}

\author{Andreas K. Bacher, Thomas B. Schr{\o}der, and Jeppe C. Dyre}\email{dyre@ruc.dk}
\affiliation{DNRF Center ``Glass and Time", IMFUFA, Department of Sciences, Roskilde University, Postbox 260, DK-4000 Roskilde, Denmark}
\date{\today}

\begin{abstract}
It has been known for a long time that many simple liquids have surprisingly similar structure as quantified, e.g., by the radial distribution function. A much more recent realization is that the dynamics are also very similar for a number of systems with quite different pair potentials. Systems with such non-trivial similarities are generally referred to as ``quasi-universal''. From the fact that the exponentially repulsive pair potential has strong virial potential-energy correlations in the low-temperature part of its thermodynamic phase diagram, we here show that a liquid is quasi-universal if its pair potential can be written approximately as a sum of exponential terms with numerically large prefactors. Based on evidence from the literature we moreover conjecture the converse, i.e., that quasi-universality only applies for systems with this property.
\end{abstract}

\maketitle

\section{Introduction}

A cornerstone of liquid-state theory is the fact that the hard-sphere (HS) model gives an excellent representation of simple liquids. This was discussed in 1959 by Bernal \cite{ber59}, and the first liquid-state computer simulations at the same time by Alder and Wainwright likewise studied the HS system \cite{ald57}. Following this the HS model has been the fundament of liquid-state theory, in particular since the 1970s when perturbation theory matured into its present form \cite{wca,bar76,han13}. The HS model is usually invoked to explain the fact that many simple liquids have very similar structure. This cannot explain, however, the more recent observation of {\it dynamic} quasi-universality \cite{ros77,kre07,cha07,bra11,gue03,pon11,sch11,lop12,lop13} or why some mathematically simple pair potentials violate quasi-universality \cite{pre05,deo06,eng07,kre09,gal12}. 

An early hint of dynamic quasi-universality was provided by Rosenfeld who in 1977 showed that the diffusion constant is an almost universal function of the excess entropy (the entropy minus that of an ideal gas at the same temperature and density) \cite{ros77}. This finding did not attract a great deal of attention at the time, but the last decade has seen renewed interest in excess-entropy scaling \cite{kre07,cha07} and, more generally, in the striking similarities of the structure and dynamics of many simple model liquids \cite{bra11,gue03,pon11,sch11,lop12,lop13}. Thus Heyes and Branka and others have documented that inverse power-law (IPL) systems with different exponents have similar structure and dynamics \cite{hey05,sco05,bra06,hey08,bra11,pon11}, Medina-Noyola and coworkers have established that quasi-universality extends to the dynamics for Newtonian as well as Brownian equations of motion \cite{gue03,lop12,lop13}, Scopigno and co-workers \cite{sco05} have documented HS-like dynamics in liquid gallium studied by quasi-elastic neutron scattering, and Liu and coworkers have suggested a mapping of a soft-sphere system's dynamics to that of the HS system \cite{sch11}. 

Further examples of similarities between apparently quite different systems, identified throughout the years but still largely unexplained, include: the {\it Young-Andersen approximate scaling principle} according to which two systems that have the same radial distribution function - even at different thermodynamic state points -- also have the same dynamics \cite{you03}; the fact that different systems have {\it similar order-parameter maps} in the sense of Debenedetti and coworkers \cite{tru00,cha07}, i.e., that when the relevant orientational order parameter is plotted against the translational one, almost identical curves result; the {\it Lindemann and other melting rules} \cite{ubb65,khr11a}; {\it freezing rules} like Andrade's finding that freezing initiates when the reduced viscosity upon cooling reaches a certain value \cite{and34} or the Hansen-Verlet rule that a liquid freezes when the maximum structure factor reaches 2.85 \cite{han69,bau83}.

In the literature the term ``quasi-universality'' refers to the general observation that different simple systems have very similar physics as regards structure and dynamics, but before proceeding we need to define the term precisely. To do this, recall first the definition of so-called reduced quantities. These are quantities made dimensionless by dividing by the appropriate combination of the following three units: the length unit is $\rho^{-1/3}$ where $\rho$ is the number density, the energy unit is $k_BT$ where $T$ is the temperature, and the time unit is $\rho^{-1/3}\sqrt{m/k_BT}$ for Newtonian dynamics where $m$ is the particle mass (for Brownian dynamics a different time unit is used \cite{IV,pon11a}). Henceforth, reduced quantities are denoted by a tilde, for instance the reduced pair distance is defined by $\tilde r \equiv \rho^{1/3}r$. 

By quasi-universality we shall mean the property of many simple model liquids that knowledge of a single quantity characterizing the structure or dynamics in reduced units is enough to determine all other reduced-unit structural and dynamic quantities to a good approximation. Note that a special case of this is Rosenfeld's observation that the excess entropy over $k_B$, a structural quantity, determines the reduced diffusion constant. Likewise, the Young-Andersen \cite{you05} approximate scaling principle is a consequence of quasi-universality as defined here -- as these authors expressed it ``... certain dynamical properties are very insensitive to large changes in the interatomic potential that leave the pair correlation function largely unchanged''.

It is straightforward to show that if quasi-universality applies, the other points mentioned  above also follow. The fundamental questions are: Which systems are quasi-universal? What causes quasi-universality? The conventional explanation of quasi-universality starts from the fact that the HS system provides an excellent reference system for structure calculations \cite{ros77,ros99,gue03,sch11,lop12,lop13}. Approximate theories of liquid dynamic properties like renormalized kinetic theory or mode-coupling theory in its simplest version predict that the dynamics is uniquely determined by the static structure factor, i.e., that structure determines dynamics. This reasoning led to the search for and discovery of dynamic quasi-universality back in 2003 (refs. \onlinecite{gue03,you03}). 

The dynamics of the HS system consists of constant-velocity free-particle motion interrupted by infinitely fast collisions, which is quite different from the continuous motion described by Newton's laws for smooth potentials. Thus it is far from physically obvious why the HS explanation of quasi-universal structure  extends to the dynamics. Moreover, while the HS explanation does account for the finding that, e.g., the Gaussian-core model \cite{sti76,pre05,zac08} violates quasi-universality -- a model without harsh repulsions except at low temperatures -- it cannot explain why some strongly repulsive or hard-core pair-potential systems violate quasi-universality. Examples of such systems are the Lennard-Jones Gaussian model \cite{eng07} and the Jagla model \cite{yan06}. Finally, the one-component plasma model is well known to be quasi-universal \cite{ros77}, but it is not intuitively clear when and why the gently varying Coulomb force can be well approximated by the harsh HS interaction.

In this paper we do not use the HS reference system. In a continuation of recent works \cite{dyr13,dyr13a,bac14} we take a different approach to quasi-universality by showing that any pair potential, which can be approximated by a sum of exponential terms with numerically large prefactors, is quasi-universal. This introduces the ``$\e$ quasi-universality class'' of pair potentials. Based on the available evidence from the literature we moreover conjecture the converse, that is, that all quasi-universal systems are in the $\e$ quasi-universality class.

\section{Results}

\subsection{The exponentially repulsive pair potential}

We study below the monatomic system described by the purely repulsive ``$\e$'' pair potential

\be\label{EXP}
v_{\esub}(r;\varepsilon,\sigma)
\,=\,\varepsilon\, e^{-r/\sigma}\,.
\ee
This was discussed already in 1932 by Born and Meyer and by Buckingham in 1938 in the context of a pair potential with an exponentially repulsive term plus an $r^{-6}$ attractive term \cite{bor32,buc38}; note also that the well-known Morse pair potential is a difference of two exponentials \cite{gir59}. The $\e$ pair potential may be justified physically as reflecting the overlap of electron wavefunctions in conjunction with the fact that bound-state wavefunctions decay exponentially in space \cite{bor32}. Although the $\e$ pair potential since 1932 has been used occasionally in computer simulations and for interpretation of experiments \cite{buc47,mas54,abr69,gup72,toda,bac14}, it never became a standard pair potential like the Lennard-Jones and Yukawa pair potentials \cite{han69,han75}. Given the mathematical simplicity of the $\e$ function this may seem surprising, but a likely explanation is that no systems in nature are believed to be well described by this purely repulsive pair potential. The present paper suggests that it may  nevertheless be a good idea to regard the $\e$ pair potential as the fundamental building block of simple liquids' physics, much like the exponential function is in the Fourier and Laplace transform theories of pure mathematics.

\begin{figure}
\centering
\includegraphics[scale = 0.39]{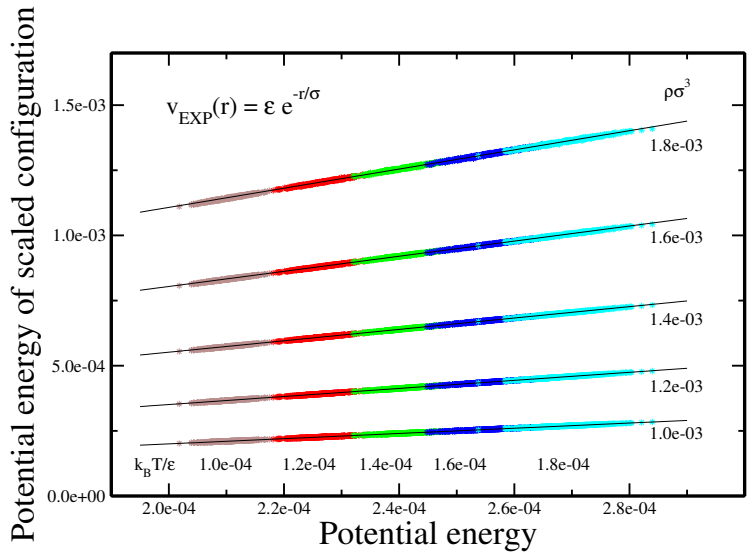}
\caption{Scatter plots of potential energies per particle of the $\e$ system. At the state point with density $\rho\sigma^3=1.0\cdot 10^{-3}$ and temperature $k_BT/\varepsilon=1.0\cdot 10^{-4}$ several configurations were selected from an equilibrium simulation and scaled uniformly to four higher densities (the light brown points); the same was done for four other temperatures at this density. The figure shows that there are very strong correlations between scaled and unscaled potential energies, with a scaling factor that only depends on the two densities involved. The lines are best fits to the green data points, that is, those generated from simulations at temperature $k_BT/\varepsilon=1.4\cdot 10^{-4}$. This figure validates the hidden-scale-invariance property of the $\e$ pair-potential system as expressed below in Eq. (\ref{prop1m}).}
\label{fig1}
\end{figure}

Our simulations focused on the low-temperature part of the thermodynamic phase diagram, i.e., where 

\be\label{req}
k_BT \ll \varepsilon\,.
\ee
The potential energy as a function of all the particle coordinates $\bR\equiv (\br_1,...,\br_N)$ is denoted by $U_{\e}(\bR)$. Figure 1 shows scatter plots of $U_{\e}(\bR)$ for several configurations $\bR$ plotted versus the potential energies of the same configurations scaled uniformly to a different density, i.e., $U_{\e}(\bR)$ plotted versus $U_{\e}(\lambda\bR)$ in which $\lambda^3$ is the ratio of the two densities in question. The figure was constructed by simulating five temperatures at the density $\rho\sigma^3=1.0\cdot 10^{-3}$ (lower line, different colors); at each of these five state points configurations were selected and scaled to four higher densities. We see that there are very strong correlations between the potential energies of scaled and unscaled configurations. 

As shown elsewhere strong correlations between scaled and unscaled potential energies are characteristic for systems that have strong correlations between their virial and potential-energy thermal equilibrium fluctuations at the relevant state points \cite{IV}. Systems with this property include not only ``atomic'' pair-potential systems like the IPL, Yukawa, and Lennard-Jones (LJ) systems, etc, but also a number of rigid molecular model systems and even the flexible LJ chain model  \cite{ing12b,vel14}. Such systems were previously referred to as ``strongly correlating'', but are now called Roskilde-simple or just Roskilde systems \cite{mal13,pra14,fle14,hen14,pie14,dyr14} which avoids confusion with strongly correlated quantum systems. 

The strength of the virial potential-energy correlations of the $\e$ system is reported in Fig. 2(a) for a large number of state points. If virial and potential energy are denoted by $W$ and $U$, respectively, and sharp brackets denote canonical constant-volume ($NVT$) averages, the color coding of the figure gives the Pearson correlation coefficient $R$ defined \cite{II} by 

\be\label{R}
R=\frac{\langle\Delta W \Delta U\rangle}{\sqrt{\langle (\Delta U)^2\rangle\langle (\Delta W)^2\rangle}}\,. 
\ee
A pragmatic definition is that a given system is Roskilde-simple at the state point in question if $R>0.9$ \cite{II}. This is the case for the $\e$ system whenever $k_BT/\varepsilon <0.1$. 

Figure 2(b) zooms in on the low-temperature part of the phase diagram where the $\e$ system has particularly strong virial potential-energy correlations. In both figures the line segment have slope $\gamma$, the slope of the so-called isomorph through the state point in question that is calculated from the expression $\gamma={\langle\Delta W \Delta U\rangle}/{\langle (\Delta U)^2\rangle}$ \cite{IV}. An isomorph is a curve in the phase diagram along which structure and dynamics to a good approximation are invariant in reduced units \cite{IV,ing12}. A system has isomorphs if and only if the system has strong virial potential-energy correlations at the relevant state points \cite{IV}. The existence of isomorphs implies that the phase diagram becomes effectively one-dimensional for many physical properties; Ref. \cite{dyr14} gives a recent review of the isomorph theory with a focus on its validation in computer simulations and experiments. The curves in Fig. 2(a) are examples of isomorphs of which the full curve is the melting isomorph \cite{IV}. Isomorphs of the $\e$ system were studied briefly in Ref. \onlinecite{bac14}.

\begin{figure}
\centering
\includegraphics[scale = 0.35]{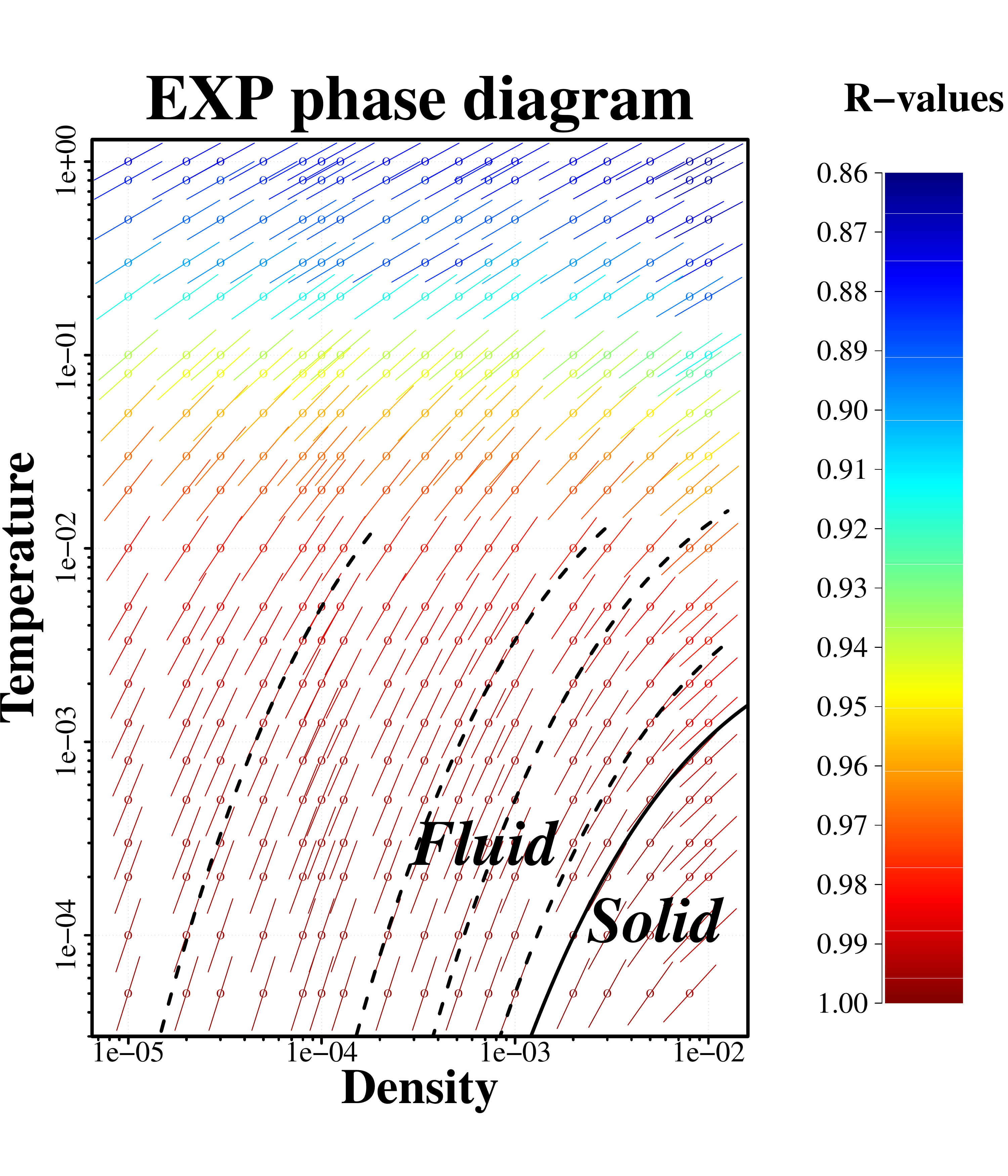}
\includegraphics[scale = 0.5]{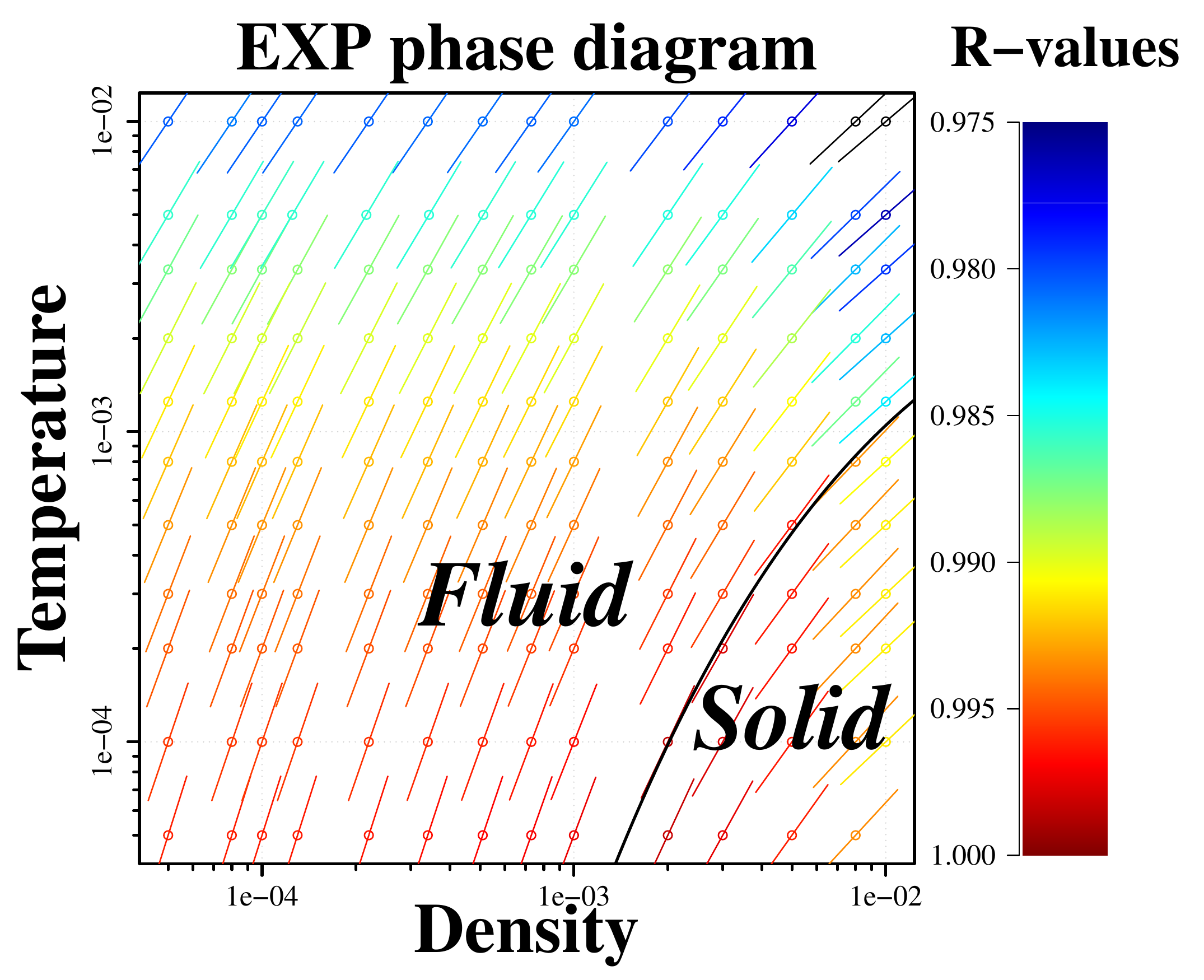}
\caption{Thermodynamic phase diagram of the $\e$ system. Density and temperature are both given in the unit system defined by the pair-potential parameters (Eq. (\ref{EXP})).
(a) gives an overview of the phase diagram and (b) zooms in on its low-temperature part; in both cases the colors indicate the value of the virial potential-energy correlation coefficient $R$ defined in Eq. (\ref{R}). The slope of each line segment is that of the isomorph through the state point in question, the curve along which structure and dynamics are (approximately) invariant in reduced units. The full curves mark the melting isomorph; this curve's width in (a) is approximately that of the coexistence region. The dashed curves mark a few other isomorphs.}
\label{fig2}
\end{figure}

\begin{figure}
\centering
\includegraphics[scale = 0.6]{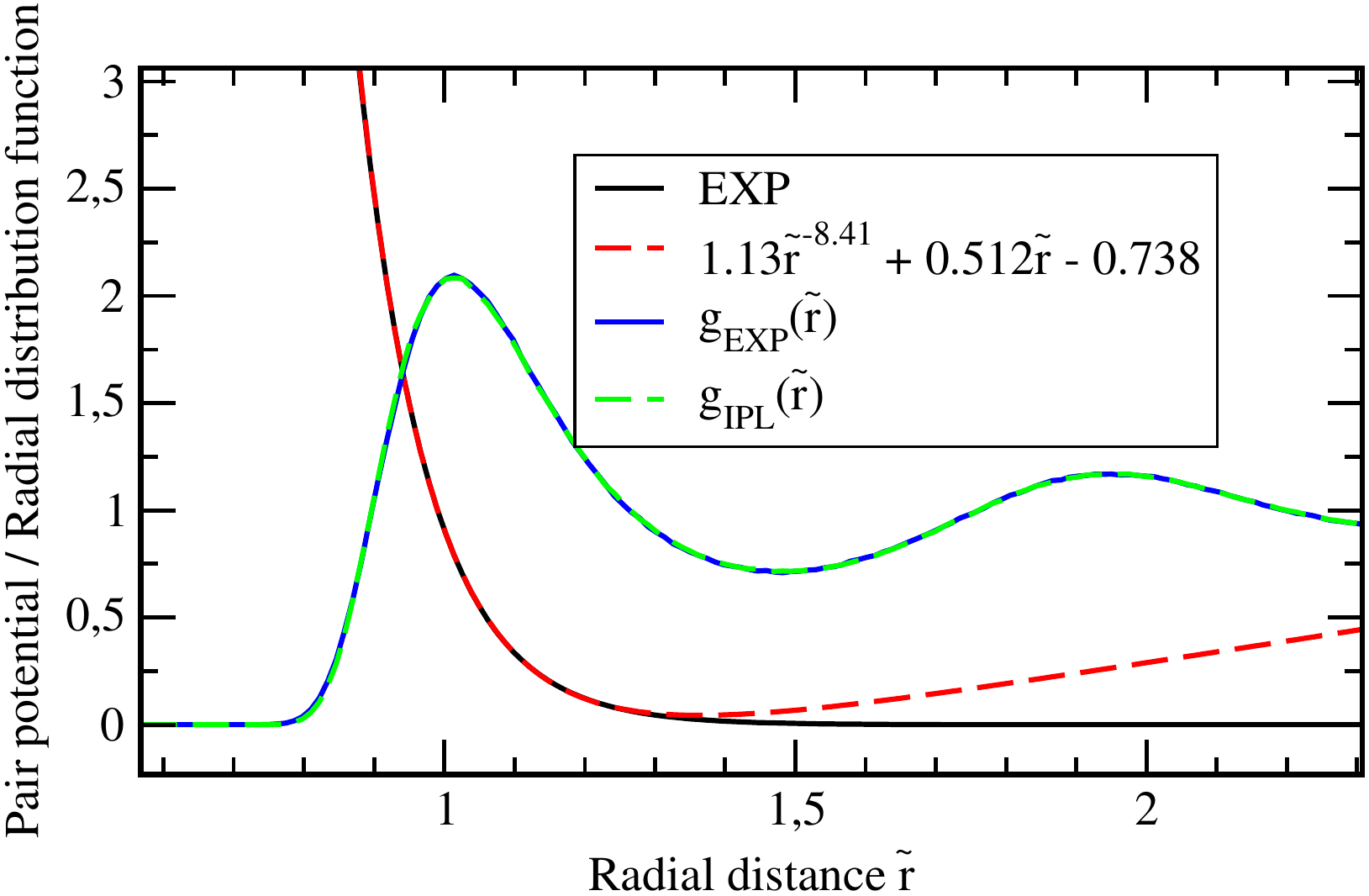}
\caption{Illustration of how the $\e$ pair potential (black curve) may be fitted by an eIPL pair potential \cite{II}. The eIPL pair potential is a sum of an inverse power-law (IPL) term and a linear term (red dashed curve); $\tilde r \equiv r\rho^{1/3}$ is the reduced pair distance and the pair potential is given in units of $k_BT$. The eIPL function was fitted to the $\e$ function over the range of radial distances for which the radial distribution function (RDF) at the state point defined by $\rho\sigma^3=1.0\cdot 10^{-3}$ and $k_BT/\varepsilon=5.0\cdot 10^{-5}$ is larger than unity. The blue and green curves are the RDFs at this state point of the $\e$ pair potential and the IPL pair potential, respectively. The $\e$ and IPL pair potentials' diffusion constants differ by less than 1\% at this state point (data not shown).}
\label{fig3}
\end{figure}

Before proceeding we address the question  why the $\e$ pair potential has strong virial potential-energy correlations. Our explanation refers to IPL pair potentials, which by Euler's theorem for homogeneous functions have 100\% virial potential-energy correlations (recall that the microscopic virial is defined generally by $W(\bR)=(-1/3)\bR\cdot\nabla U(\bR)$ \cite{han13}). Figure \ref{fig3} refers to the state point $\rho\sigma^3=1.0\cdot 10^{-3}$, $k_BT/\varepsilon=5.0\cdot 10^{-5}$. The figure shows that the $\e$ pair potential (black curve) may be fitted very well over the entire first coordination shell by an extended IPL (``eIPL'') function, which is defined as an IPL term plus a linear term (the red dashed line). As shown in Ref. \cite{II} the linear term contributes little to neither the virial nor the potential-energy constant-volume fluctuations because the sum of nearest-neighbor distances is almost constant. This is because if one particle is moved, some nearest-neighbor distances decrease and others increase, resulting in almost no change in the sum of the nearest-neighbor distances (this argument is exact in one dimension). As a consequence, the $WU$ fluctuations are dominated by the IPL term and thus strongly correlating. This explains why the $\e$ pair potential has very strong $WU$ correlations. To confirm the equivalence between the $\e$ pair potential and the IPL pair potential ($1.13\,\tilde r^{-8.41}$), we performed a simulation with the IPL pair potential at the same state point. As predicted \cite{II,ped10} there is good agreement between the two systems' structure (blue and green curves) and dynamics (not shown), confirming that the linear term of the eIPL pair potential fitting the $\e$ pair potential, $0.512\,\tilde r-0.738$, is not important for the physics.

\subsection{Derivation of quasi-universality}

We proceed to show that the strong virial potential-energy correlation property of the $\e$ system implies quasi-universality for a large class of monatomic systems. As shown recently \cite{dyr13a}, any Roskilde-simple system, that is, system with strong virial potential-energy correlations and isomorphs, is characterized by ``hidden scale invariance'' in the sense that two functions of density exist, $h(\rho)$ and $g(\rho)$, such that the potential-energy function $U(\bR)$ can be expressed as follows 

\be\label{prop1m}
U(\bR)\cong h(\rho)\tU(\tbR)+g(\rho)\,.
\ee
Here $\tU(\tbR)$ is a dimensionless, state-point-independent function of the reduced dimensionless configuration vector $\tbR\equiv\rho^{1/3}\bR$. In particular, Eq. (\ref{prop1m}) implies the strong correlations between the potential energies of scaled and unscaled configurations documented in Fig. 1: Consider two configurations $\bR_1$ and $\bR_2$ at density $\rho_1$ and $\rho_2$, respectively, with the same reduced coordinates, i.e., obeying $\rho_1^{1/3}\bR_1=\rho_2^{1/3}\bR_2$. By elimination of $\tU(\tbR)$ in Eq. (\ref{prop1m}) we get $U(\bR_2)\cong (h(\rho_2)/h(\rho_1))U(\bR_1)+g(\rho_2)-g(\rho_1)h(\rho_2)/h(\rho_1)$. Thus any two configurations with the same reduced coordinates have potential energies that are approximately linear functions of each other with constants that only depend on the two densities in question, compare Fig. 1.

According to Eq. (\ref{prop1m}) a change of density implies that the potential-energy landscape to a good approximation undergoes a linear, affine transformation. Moreover, Eq. (\ref{prop1m}) implies invariant structure and dynamics in reduced units along the curves defined by constant $h(\rho)/k_BT$, the equation identifying the isomorphs \cite{dyr13a,dyr14,IV}: along any curve defined by $h(\rho)/k_BT=C$ the reduced force $\tbF\equiv -\tilde\nabla U(\bR)/k_BT$ is given by $\tbF=-C\,\tilde\nabla\tU(\tbR)$, which implies that $\tbF$ is a certain function of the reduced coordinates, i.e., $\tbF=\tbF(\tbR)$. Since Newton's second law in reduced coordinates is $d^2\tbR/d\tilde{t}^2=\tbF$ (the reduced mass is unity), it follows that the particles move in the same way at different isomorphic state points -- except for the trivial linear uniform scalings of space and time involved in transforming back to real units.

Since the $\e$ pair potential has strong virial potential-energy correlations, Eq. (\ref{prop1m}) implies that functions $h_{\esub}(\rho)$, $g_{\esub}(\rho)$, and $\tUe(\tbR)$ exist such that 

\be\label{hsiq}
U_{\e}(\bR)\cong h_{\esub}(\rho)\tUe(\tbR)+g_{\esub}(\rho)\,.
\ee
The functions $h_{\esub}(\rho)$ and $g_{\esub}(\rho)$ in Eq. (\ref{hsiq}) both have dimension energy. They can therefore be written $h_{\esub}(\rho)=\varepsilon\,\thf_{\esub}(\rho\sigma^3)$ and $g_{\esub}(\rho)=\varepsilon\,\tgf_{\esub}(\rho\sigma^3)$ in which $\thf_{\esub}$ and $\tgf_{\esub}$ are dimensionless functions that only depend on the dimensionless density $\rho\sigma^3$. 

Consider now a pair potential $v(r)$ that can be expressed as follows 

\be\label{vr}
v(r)=\int_0^\infty f(\sigma) e^{-r/\sigma} d\sigma\,.
\ee
The system's potential energy is the sum of the individual pair-potential contributions with the same weights as in Eq. (\ref{vr}). Therefore, if $U(\bR)$ is the potential energy and we define $h(\rho)\equiv\int_0^\infty f(\sigma) \thf_{\esub}(\rho\sigma^3)d\sigma$ and $g(\rho)\equiv \int_0^\infty f(\sigma) \tgf_{\esub}(\rho\sigma^3)d\sigma$, it follows from Eq. (\ref{hsiq}) that 

\be\label{veq}
U(\bR)\cong h(\rho)\tUe(\tbR)+g(\rho)\,.
\ee
Because the reduced-unit physics is encoded in the function $\tUe(\tbR)$ via Newton's equation $d^2\tbR/d\tilde{t}^2=-C\,\tilde\nabla\tUe(\tbR)$ where $C=h(\rho)/k_BT$ identifes the isomorph through the state point in question, Eq. (\ref{veq}) implies identical structure and dynamics to a good approximation for systems obeying Eq. (\ref{vr}). This argument would be exact if the $\e$ pair potential had 100\% virial potential-energy correlations. This is not the case, though, and the approximation Eq. (\ref{veq}) is only useful when it primarily involves $\e$ functions from the low-temperature part of the phase diagram where strong virial potential-energy correlations and thus Eq. (\ref{prop1m}) apply.

We have shown that a pair potential $v(r)$ is quasi-universal if it can be written as a sum of low-temperature $\e$ pair potentials. To translate this into an operational criterion, note the following. If the integral in Eq. (\ref{vr}) is discretized into a finite sum and expressed in terms of the reduced pair potential $\tilde v \equiv v/k_BT$ regarded as a function of the reduced pair distance ($\tilde r \equiv\rho^{1/3}r$), the condition for quasi-universality is that

\be\label{app}
\tilde v(\tilde r)\cong \sum_j \Lambda_j e^{-u_j\tilde r}\,\,,\,\, \lvert\Lambda_j\rvert\gg 1\,. 
\ee
It is understood that the ``wavevectors'' $u_j$ are not so closely spaced that large positive and negative neighboring terms may almost cancel one another. Several points should be noted: 1) Equation (\ref{app}) is state-point dependent because the function $\tilde v(\tilde r)$ varies with state point; 2) a continuous integral of $\e$ functions does not automatically obey Eq. (\ref{app}) -- it is necessary that the integral can be approximated by a {\it finite} sum of $\e$ terms, each with a numerically large prefactor; 3) a sum or product of two pair potentials obeying Eq. (\ref{app}) with all $\Lambda_j >0$ gives a function that also obeys this equation.

\subsection{Important examples}

Consider first the IPL pair potential $v_n(r)\equiv\varepsilon \left(r/{\sigma}\right)^{-n}$. In terms of the reduced radius $\tr$, the reduced IPL pair potential is given by
$\tv_n(\tr)\equiv v_n(r)/k_BT = \Gamma_n\tr^{-n}$ in which $\Gamma_n\equiv (\rho\sigma^3)^{n/3}\,\varepsilon/k_BT$. The mathematical identity $\int_0^\infty x^{n-1}\exp(-x)dx=(n-1)!$ implies that 
$\tv_n(\tr)=[\Gamma_n/(n-1)!]\int_0^\infty u^{n-1}\exp(-u\tr)du$. Discretizing the integral leads to
$\tv_n(\tr)\cong[\Gamma_n/(n-1)!]\,\Delta u\, \sum_{j=0}^\infty ((j+1/2)\Delta u)^{n-1}\exp(-(j+1/2)\Delta u\tr)$. Writing $((j+1/2)\Delta u)^{n-1}=\exp[(n-1)\ln((j+1/2)\Delta u)]$, by differentiation with respect to $j$ it is easy to see that the dominant contributions to the sum come from the terms with $(n-1)/(j+1/2)\simeq\Delta u\tr$. Thus for typical nearest-neighbor distances ($\tr\simeq 1$) the terms with $(j+1/2)\Delta u\simeq n-1$ are the most important ones. For these values of $j$ the prefactor of the exponential in the above sum is roughly $\Gamma_n\,\Delta u\, (n-1)^{n-1}/(n-1)!$. The largest realistic discretization step $\Delta u$ is of order unity, so we conclude that for values of $n$ larger than three or four Eq. (\ref{app}) is obeyed unless $\Gamma_n$ is very small, a condition that applies for the state points that have typically been studied \cite{hey05,sco05,bra06,hey08,bra11}. The case of a Coulomb repulsive system ($n=1$) is discussed in the next section.

As a consequence of the above, at most state points the potential energy of the IPL system, $U_n(\bR)$, can be written

\be\label{hsipl}
U_{n}(\bR)\cong h_n(\rho)\tUe(\tbR)+g_n(\rho)\,.
\ee
Since $U_n(\bR)\propto \rho^{n/3}$ for the density variation induced by a uniform scaling of a configuration $\bR$, i.e., keeping $\tbR$ constant, one has $h_n(\rho)\propto\rho^{n/3}$ and $g_n(\rho)\propto\rho^{n/3}$. This means that two numbers $\alpha_n$ and $\beta_n$ exist such that $U_{n}(\bR)\cong\varepsilon\left[\alpha_n(\rho\sigma^3)^{n/3}\tUe(\tbR)+\beta_n(\rho\sigma^3)^{n/3}\right]$. For a general pair potential of the form $v(r)=\varepsilon\sum_nv_n(r/\sigma)^{-n}$, by a linear combination of Eq. (\ref{hsipl}) we arrive at Eq. (\ref{veq}) in which $h(\rho)=\varepsilon\sum_n v_n\alpha_n(\rho\sigma^3)^{n/3}$ and $g(\rho)=\varepsilon\sum_n v_n\beta_n(\rho\sigma^3)^{n/3}$. 

A well-known case is the Lennard-Jones (LJ) pair potential $v_{\rm LJ}(r)=4\varepsilon[(r/\sigma)^{-12}-(r/\sigma)^{-6}]$. Consider the LJ liquid state point given by $\rho\sigma^3=1$, $k_BT=2\varepsilon$. In terms of the above-defined reduced IPL functions, it is easy to see that this is of the form Eq. (\ref{app}), implying quasi-universality of the LJ liquid at this state point. 

An application of the $\e$ theory of quasi-universality is the intriguing ``additivity of melting temperatures'' \cite{ros76} according to which if two systems have melting temperatures that as functions of density are denoted by $T_{{\rm m},1}(\rho)$ and $T_{{\rm m},2}(\rho)$, the melting temperature of the system with the sum potential energy is $T_{{\rm m},1}(\rho)+T_{{\rm m},2}(\rho)$. This property follows from quasi-universality because the dynamics of all three systems are controlled by the same function $\tUe(\tbR)$. For the $\e$ system melting initiates when this function's average upon heating reaches a certain value, and the same must apply for all quasi-universal systems. In particular, since for an IPL system one has $T_{{\rm m}}(\rho)\propto\rho^{n/3}$ \cite{sti75}, the melting temperature of the LJ system varies with density according to the expression $T_{\rm m}(\rho)= A\rho^4-B\rho^2$ (refs. \onlinecite{ros76,khr11}).

\section{Discussion}

Denominating the systems that obey Eq. (\ref{app}) collectively as the {\it $\e$ quasi-universality class}, we have shown that all systems in this class are quasi-universal in the sense of this paper: if a single reduced-unit structural or dynamic quantity is known, all other reduced-unit structural or dynamic quantities are known to a good approximation. This is because these are all encoded in the function $\tUe(\tbR)$, and any reduced-unit quantity characterizing structure or dynamics identifies the constant $C=h(\rho)/k_BT$ of the reduced-unit version of Newton's second law $d^2\tbR/d\tilde{t}^2=-C\,\tilde\nabla\tUe(\tbR)$.

The obvious question is whether all quasi-universal systems are in the $\e$ quasi-universality class. We cannot prove this, but conjecture it is the case based on the available evidence in the literature for the following systems: 

The Jagla pair potential. This is a HS potential plus a finite-width potential well defined by two terms that are linear in $r$ \cite{yan06}. This pair potential is not in the $\e$ quasi-universality class -- it cannot be approximated as a sum of exponentials because its Laplace transform only has a pole at zero. Indeed, the Jagla pair potential reproduces water's anomalous density maximum and has a liquid-liquid critical point \cite{gal12}, properties which are both inconsistent with quasi-universality. 

The Gaussian core model (GCM). This is a Gaussian centered at $r=0$. For this system there is a reentrant BCC phase above the triple point \cite{pre05} and the transport coefficients have a non-monotonous density dependence at constant temperature \cite{kre09}. These observations both contradict quasi-universality. Consistent with the above conjecture, a representation of the form $\exp(-r^2/2\sigma^2)=\int_0^\infty \phi(u)e^{-ur}du$ does not exist because the Laplace transform of a Gaussian has no poles; thus the GCM system is not in the $\e$ quasi-universality class. At low temperatures and low densities the GCM pair potential can be approximated well by an exponential, however, and in this part of the phase diagram the GCM system is indeed quasi-universal \cite{pre05,pon11}. 

The LJ Gaussian (LJG) model. This pair potential is arrived at by adding a negative, displaced Gaussian to the LJ pair potential \cite{eng07}. Like the GCM it cannot be written as a sum of exponentials. The LJG system has thermodynamic and dynamic anomalies \cite{deo06} and ``a surprising variety of crystals" \cite{eng07}, both of which are observations that violate quasi-universality. 

The one-component plasma (OCP). This is the common name for the single-charge Coulomb system. The Coulomb potential is too long ranged for a thermodynamic limit to exist for the free energy per particle unless a uniform charge-compensating background is introduced \cite{han13,han75}. Nevertheless, any finite OCP system is well defined and amenable to computer simulation -- in fact the OCP system was an important example in Rosenfeld's original paper on excess-entropy scaling \cite{ros77}. The OCP system is the $n=1$ case of the above discussed IPL pair potential; it is quasi-universal in the dense fluid case, i.e., whenever $\Gamma_1 \gg 1$ (writing $\tilde v_1\equiv\Gamma_1/\tilde r$). Violations of quasi-universality are indeed known to gradually appear when $\Gamma_1$ goes below 50 \cite{dal06}. 

The Yukawa pair potential. This is given by the expression $v(r)=\varepsilon (\sigma/r)\exp(-r/\sigma)$. Since $v(r)/\varepsilon=\int_0^\infty\exp(-r(u+1)/\sigma)du$ it is easy to see that this pair potential is in the $\e$ quasi-universality class whenever $k_BT\ll\varepsilon$. This is consistent with known properties \cite{ros01}. 

This paper has argued that one may replace the HS by the $\e$ system as the generic simple pair-potential system from which quasi-universality is derived. We do not wish to suggest that the HS system's role in liquid-state physics is entirely undeserved, however, because this model is still uniquely simple and mathematically beautiful. Nevertheless, by using the $\e$ pair potential as the fundamental building block, a number of advantages are obtained. First of all, smoothness is ensured. Secondly, if the conjecture that all quasi-universal functions are in the $\e$ quasi-universality class is confirmed, we now have a mathematically precise characterization of all quasi-universal pair potentials (Eq. (\ref{app})). Thirdly, the present explanation of quasi-universality gives a natural explanation of dynamic quasi-universality. Finally, the role of the HS pair potential in liquid-state physics is clarified: quasi-universality is not {\it caused} by a given system's similarity to the HS system -- rather, quasi-universality {\it implies} similarity between the properties of many systems and those of the HS system simply because the latter is in the $\e$ quasi-universality class by being the $n\rightarrow\infty$ limit of an IPL system. -- Note that while this paper focused on quasi-universality for liquid models, the crystalline phases of these models are likewise quasi-universal, for instance by having quasi-universal radial distribution functions, phonon spectra, and vacancy jump dynamics \cite{alb14}.

In continuing work it will be interesting to investigate the consequences of the $\e$ approach for the fact that quasi-universality appears to apply beyond the framework of thermal equilibrium, e.g., for the jamming transition in which the HS model is presently used as the generic model \cite{ler12}. It is an open question whether replacing the HS model by the $\e$ model in thermodynamic perturbation theory would have advantages \cite{han13}. Another open question is to which extent mixtures are quasi-universal \cite{ler14}. Finally, recent results from the works of Truskett and collaborators \cite{pon11} show that classical Rosenfeld excess-entropy scaling may be modified into more general excess-entropy scalings, and it would be interesting to investigate whether the present approach can somehow be extended to account for this.

\section{Methods}

A system of $N = 1,000$ particles interacting via the $\e$ pair potential was simulated using standard Nose-Hoover $NVT$ simulations with a time step of $0.0025$ and thermostat relaxation time of $0.2$ (LJ units). A shifted-forces cutoff at $r=2\,\rho^{-1/3}$ was used for densities below $1.0\cdot 10^{-3}$, at larger densities the cutoff was $4\,\rho^{-1/3}$. At each state point the simulations involved $10,000,000$ time steps after equilibration. For temperatures below $2.0\,\cdot 10^{-3}$ the simulations were initiated from a state of $2,000$ particles placed in a body-centered cubic crystal structure. The melting isomorph was determined by the interface pinning method \cite{ped13} -- the $NPT$ simulations involved here were made using LAMMPS \cite{LAMMPS} (http://lammps.s andia.gov) with shifted-potential cutoffs ranging from $2.5\,\rho^{-1/3}$ to $4.0\,\rho^{-1/3}$ for a system of $2,560$ particles; the time step was $0.005$ and the relaxation time was $0.4$ for the thermostat and $0.8$ for the barostat.\\

\vspace{1cm}

{\bf Acknowledgments}\\
We are indebted to Ulf Pedersen for help in determining the freezing and melting isomorphs. The center for viscous liquid dynamics ``Glass and Time'' is sponsored by the Danish National Research Foundation's grant DNRF61.\\\newline\vspace{1cm}

{\bf Author contributions}\\
A.K.B. and T.B.S. performed the simulations and carried out the data analysis, J.C.D. conceived the project and wrote the paper.\\\newline

\end{document}